\documentclass[12pt]{article}

\usepackage{amssymb}

\begin{document}

\begin{titlepage}

\begin{flushright}
arXiv:1603.04491
\end{flushright}
\vskip 2.5cm

\begin{center}
{\Large \bf Cerenkov-Like Emission of Pions by\\
Photons in a Lorentz-Violating Theory}
\end{center}

\vspace{1ex}

\begin{center}
{\large Brett Altschul\footnote{{\tt baltschu@physics.sc.edu}}}

\vspace{5mm}
{\sl Department of Physics and Astronomy} \\
{\sl University of South Carolina} \\
{\sl Columbia, SC 29208} \\
\end{center}

\vspace{2.5ex}

\medskip

\centerline {\bf Abstract}

\bigskip

In the presence of Lorentz violation, the Cerenkov-like process
$\gamma\rightarrow\gamma+\pi^{0}$ may become allowed for sufficiently
energetic photons. Photons above the threshold would lose energy
rapidly through pion emission. The fact that
propagating photons with energies of
up to 80 TeV survive to be observed on Earth allows us to place a
one-sided constraint on an isotropic Lorentz violating parameter at the
$7\times 10^{-13}$ level; this is more than an order of magnitude better than
the best previous result.

\bigskip

\end{titlepage}

\newpage

One interesting possibility for new physics beyond the standard model
is that there might be violations of Lorentz and CPT symmetries. So far,
these are strictly theoretical possibilities; there has been no compelling
evidence to suggest that these symmetries are not truly exact in nature.
However, if these symmetries turn out not to hold exactly, that would
certainly be a discovery of great importance. The discovery would open up
a window onto a completely new regime of physics.

The fundamental physical theories that we currently understand---the
standard model and general relativity---are both fully invariant under
Lorentz symmetry and CPT. However, the machinery of effective field theory
can be used to describe modified versions of these theories that do not
respect the symmetries exactly.
The most general effective field theory that is used to describe such
symmetry violations is known as the standard model extension (SME).
The SME action contains all the possible translation-invariant but
Lorentz-violating operators that could be
constructed out of standard boson and fermion fields.
The operators that have residual tensor indices violate Lorentz symmetry,
and they may break CPT symmetry as well.
The coupling constants of the theory are vector and tensor valued; their
Lorentz indices are contracted
with the free indices found on the field operators~\cite{ref-kost1,ref-kost2}.
If the origin of the Lorentz violation is spontaneous symmetry breaking,
these tensors, which represent preferred directions in spacetime, are
related to the vacuum expectation values of dynamical vector and
tensor fields.

One of the virtues of the SME is its generality; it contains operators of
arbitrarily high mass dimension. However, a
specific restricted version of the SME has become the standard framework
for parameterizing the results of most empirical Lorentz and CPT tests.
The restricted version is known as the minimal SME, and it
contains only local, gauge-invariant operators of renormalizable mass
dimension. With these restrictions in place, the minimal SME contains only a
finite number of undetermined parameters, and radiative corrections are
systematically calculable~\cite{ref-kost3,ref-collad-3,ref-ferrero3}.
This makes the minimal SME framework well suited
for comparing the results of different experiments, and experiments in many
different areas of physics have been used to constrain the minimal SME
coefficients. The best current constraints are summarized in~\cite{ref-tables}.

Many of the tightest constraints on the Lorentz-violating operators of the
SME come from analyses of astrophysical data. Two particular things are
available in extraterrestrial environments that can make astrophysical
tests of Lorentz symmetry
extremely sensitive: very large distances and very high energies. Observations
of photons that have traversed cosmological distances can
be exquisitely sensitive to the tiniest changes in electromagnetic wave
propagation. The strongest constraints on other effects come from looking at the
emissions from extremely energetic astronomical sources, in which
individual particles may have energies at up to the PeV scale.

By studying the photons emitted by such sources, we may learn a
great deal about the energy-momentum relations not just for the
photons themselves, but other particles as well. The observation of
a photon with a TeV-scale energy coming from an extraterrestrial source
can tell us things in two different ways. If the process that is
responsible for the photon emission at the source is understood, 
the energies of the emitted photons can reveal important information
about the energy-momentum relations for the other particles involved
in the emission process. The typical processes that produce
ultra-high-energy $\gamma$-rays are inverse Compton
scattering, $e^{-}+\gamma\rightarrow e^{-}+\gamma$ (in which a
low-energy photon is upscattered by an extremely energetic electron);
and neutral pion decay, $\pi^{0}\rightarrow\gamma+\gamma$. When
we observe photons that we know to originate from one or the other
of these processes, we learn a substantial amount about the
energy-momentum relations for the massive particles
involved. This has made it possible to place strong constraints on 
Lorentz violation in the electron sector of the
SME~\cite{ref-jacobson1,ref-altschul6,ref-altschul15} and
somewhat weaker constraints in the pion sector~\cite{ref-altschul16}.

Moreover, observations of cosmic ray $\gamma$-rays can also tell us
interesting things about Lorentz violation in the electron and pion
sectors, even when the processes by which the photons were produced are
unknown. The reason is that, while photons in the Lorentz-invariant
standard model are absolutely stable in vacuum, sufficiently energetic
photons may actually decay if there is Lorentz violation present.
For a photon from a distant source to survive long enough to reach
Earth, photon decay processes such as pair creation,
$\gamma\rightarrow e^{-}+e^{+}$, or Cerenkov-like pion emission,
$\gamma\rightarrow\gamma+\pi^{0}$, must be forbidden (or exceedingly
slow) at the observed photon energy. This provides a complementary way
of constraining Lorentz violation outside the photon sector using
$\gamma$-ray observations.

Previously, the survival of TeV photons has been used to place
complementary constraints on electron-sector Lorentz violation. The fact
that these photons live long enough to be seen ensures that
$\gamma\rightarrow e^{-}+e^{+}$ is not occurring~\cite{ref-stecker}.
Of course, an
electron and a positron are not the only particle pair that
might be produced in such a reaction. The fact that photons do not
decay via $\gamma\rightarrow\pi^{+}+\pi^{-}$ allows us to place bounds
on the SME coefficients for charged pions~\cite{ref-altschul14}.
These bounds are weaker than
the electron-sector bounds,
because the typical strength of a bound derived from
a high-energy astrophysical process is $\sim m^{2}/E^{2}$, where
$m$ is the mass of the heaviest particle involved in the process and
$E$ is the energy scale of the process. Yet while this means that the
constraints in the charged pion sector are orders of magnitude worse
that the equivalent constraints in the electron sector, the pion bounds
are still important. The reason is that bounds on pion Lorentz violation
are typically rather difficult to obtain. In fact, it is generally true
that the constraints on Lorentz violation for unstable species are
frequently quite weak, and the goal of this paper will be to improve
the constraints on Lorentz violation for the even shorter lived neutral
pion.

Lorentz violation in the SME is described by tensor-valued coefficients
contracted with tensor operators constructed out of the particle fields.
The minimal SME Lagrange density for the free neutral pion field is
\begin{equation}
{\cal L}_{\pi}=\frac{1}{2}(\partial^{\mu}\pi^{0})(\partial_{\mu}\pi^{0})+\frac{1}{2}
k^{\mu\nu}(\partial_{\mu}\pi^{0})(\partial_{\nu}\pi^{0})
-\frac{m_{\pi}^{2}}{2}(\pi^{0})^{2}.
\end{equation}
The Lorentz violation in the pure pion sector is governed by the
nine small coefficients contained in the traceless, symmetric $k^{\mu\nu}$
tensor. These coefficients, which appear in the action for the composite
pion field, must
ultimately be related to the SME coefficients for the more
fundamental quark and gluon fields.

The pions are coupled to photons, and the Lagrange
density for the pure photon sector and the pion-photon coupling is
\begin{equation}
\label{eq-L-A}
{\cal L}_{A}=-\frac{1}{4}F^{\mu\nu}F_{\mu\nu}
-\frac{1}{4}k_{F}^{\mu\nu\rho\sigma}F_{\mu\nu}F_{\rho\sigma}
+\frac{1}{2}k_{AF}^{\mu}\epsilon_{\mu\nu\rho\sigma}F^{\nu\rho}
A^{\sigma}
-g\pi^{0}\epsilon^{\mu\nu\rho\sigma}F_{\mu\nu}F_{\rho\sigma}.
\end{equation}
The electromagnetic action may also contain Lorentz violation,
governed by the $k_F$ and $k_{AF}$ coefficients. However, we shall consider only
theories with vanishing $k_{F}=0$
and $k_{AF}=0$. Photon Lorentz violation is comparatively easy
to constrain; some of the coefficients have been bounded extremely tightly
using cosmological searches for photon
birefringence~\cite{ref-carroll1,ref-kost21,ref-mewes5}, while the remainder
have been bounded at a far less stringent (but still quite respectable)
level using direct Michelson-Morley tests of the isotropy of photon
propagation~\cite{ref-nagel}. In contrast, bounds on the SME
coefficients for unstable
massive particles like the $\pi^{0}$ are much weaker. So we shall focus
on the sensitivity of various observables to the less well constrained
pion $k^{\mu\nu}$ parameters.

The key to using photon observations to constrain $\pi^{0}$ Lorentz
violation is, of course, the pion-photon coupling. The coupling,
which is normally responsible for the $\pi^{0}\rightarrow\gamma+\gamma$
decay, is dominated by the chiral anomaly
$g\approx\frac{N_{c}e^{2}}{96\pi^{2}f_{\pi}}$, where $N_{c}=3$ is the
number of colors and $f_{\pi}$ the pion decay constant. In a
Lorentz-violating theory, the Cerenkov-like process
$\gamma\rightarrow\gamma+\pi^{0}$, which is ordinarily forbidden by
energy-momentum conservation, may become allowed above a certain energy
threshold.

The Lorentz-violating dispersion relation for an ultrarelativistic
particle with mass $m$ typically takes the form
\begin{equation}
\label{eq-Eofp}
E=\sqrt{m^{2}+[1+2\delta(\hat{p})]\vec{p}\,^{2}}.
\end{equation}
The parameter $\delta(\hat{p})$ determines the maximum achievable
velocity (MAV) for the particle type in question. The MAV
$1+\delta$ depends on
the direction $\hat{p}$ of the momentum (and for a fermionic
particles, $\delta$ would also depend on the helicity).
In general, $\delta$ is a function of the dimensionless
coefficients multiplying the dimension-four operators in the
relevant sector of the minimal SME. There may be additional terms
in the energy-momentum relation
derived from dimension-three operators, but their importance
diminishes with increasing energies, and so they have relatively
little impact on highly relativistic processes.

For the $\pi^{0}$ the MAV is set by
\begin{equation}
\delta(\hat{p})=-\frac{1}{2}\left[k_{00}+k_{(0j)}\hat{p}_{j}+k_{jk}
\hat{p}_{j}\hat{p}_{k}\right],
\end{equation}
where $k_{(0j)}=k_{0j}+k_{j0}$. With observations of photons coming
from a sufficient number of different directions, it could be possible
to place separate bounds on all nine of the $k^{\mu\nu}$ parameters.
However, we shall restrict our attention to the case of an isotropic
theory with constant $\delta=-\frac{2}{3}k_{00}$ (taking into account
the tracelessness of $k^{\mu\nu}$).

We shall now look in detail at how the $\gamma\rightarrow\gamma+\pi^{0}$
process behaves in the presence of a nonzero pion $\delta$. The most
obvious fact is that the process can only occur if $\delta<0$. In the
standard theory, the photon carries insufficient energy to create the
pion. However, with a negative $\delta$, the pion will have less energy
than it would in the standard theory at the same momentum. Although a
low-energy photon will still not have sufficient energy to emit a pion,
the Cerenkov-like emission will become allowed above a certain photon
threshold energy.

The kinematics of the pion emission are fairly straightforward. The
initial photon has momentum $q^{\mu}=(E,0,0,E)$. The outgoing photon
loses energy and is deflected by an angle $\theta$; it carries
momentum $q'^{\mu}=(E',E'\sin\theta,0,E'\cos\theta)$. This leaves the
pion carrying momentum $p^{\mu}=(E_{\pi},-E'\sin\theta,0,E-E'\cos\theta)$.
The pion energy-momentum relation (\ref{eq-Eofp}) dictates that
\begin{equation}
(E-E')^{2}=E_{\pi}^{2}=m_{\pi}^{2}+(E^{2}+E'^{2}-2EE'\cos\theta)(1+2\delta),
\end{equation}
so the relationship among the various quantities can be expressed
\begin{equation}
\label{eq-angles}
\sin^{2}(\theta/2)=-(1-2\delta)\frac{m_{\pi}^{2}}{4EE'}-
\delta\frac{(E-E')^{2}}{2EE'}.
\end{equation}

The pion emission process can only occur above the threshold energy $E_{T}=\frac{m_{\pi}}
{\sqrt{-2\delta}}$. If $\delta$ is negative, then for sufficiently
large $E>E_{T}$, the Cerenkov-like emission becomes possible. In the
threshold configuration, the pion carries away all the energy, and
the outgoing photon energy $E'$ is zero. Well above threshold, a typical
decay will have the pion and photon each carrying off a substantial
fraction of the energy, since decays with comparable $E'$ and
$E_{\pi}$ have the largest available phase space.  In other words, the
outgoing particles are usually beamed into a narrow pencil of angles.

However, decays with large photon deflection angles are still possible,
even though they are strongly disfavored. In a decay with a vanishing
deflection angle $\theta$, the photon carries an energy
$E'=E-\sqrt{E_{T}^{2}+m_{\pi}^{2}}$, which is obviously the majority of
the energy if the initial energy $E$ is well above the threshold.
At the other extreme, a photon with energy $E'=m_{\pi}^{2}/E$ will
recoil back with angle $\theta=\pi$ after the decay.

The matrix element for the process is quite straightforward to calculate.
The Feynman rule for the pion-photon-photon vertex, derived from the
Lagrange density~(\ref{eq-L-A}), has a factor
$8ig\epsilon^{\alpha\mu\beta\nu}q_{1\alpha}q_{2\beta}$, where
$q_{1}$ and $q_{2}$ are the photon momenta
directed into the vertex. The matrix
element for the Cerenkov-like process is then
\begin{equation}
i{\cal M}=8ig\epsilon^{\alpha\mu\beta\nu}q_{\alpha}(-q'_{\beta})
\varepsilon_{\mu}(q)\varepsilon^{*}_{\nu}(q'),
\end{equation}
where the $\varepsilon$ are the appropriate polarization vectors.
So the matrix element squared (summed over final polarizations)
for the process is
\begin{eqnarray}
\sum|{\cal M}|^{2} & = & 32g^{2}\epsilon^{\alpha\mu\beta\nu}
\epsilon^{\gamma\rho\delta\sigma}q_{\alpha}q_{\gamma}q'_{\beta}
q'_{\delta}g_{\mu\rho}g_{\nu\sigma} \\
& = & -64g^{2}[q^{2}q'^{2}-(q\cdot q')^{2}] \\
& = & 256g^{2}E^{2}E'^{2}\sin^{4}\theta.
\end{eqnarray}

The fact that the matrix element ${\cal M}$ is proportional to
$\theta^{2}$ for small deflection angles has a relatively
straightforward explanation,
tied to the involvement of the totally antisymmetric Levi-Civita
$\epsilon$-tensor. The initial photon has four-momentum
$q^{\mu}=(E,0,0,E)$, and after the emission, the outgoing
photon has $q'^{\mu}=(E',E'\sin\theta,0,E'\cos\theta)$. For small
values of $\theta$, $E'\approx E-p_{3}$, and so
\begin{equation}
\label{eq-qprime}
q'^{\mu}\approx(E-p_{3},0,0,E-p_{3})+(0,-p_{1},0,0).
\end{equation}
The vectors $q$ and $q'$ are
both contracted with an $\epsilon$-tensor. The first
term on the right-hand side of (\ref{eq-qprime}) therefore makes no
contribution, because it is proportional to $q^{\mu}$.
This leaves the whole matrix element proportional to
$-p_{1}=E'\sin\theta$.

This accounts for the presence of one factor of $\sin\theta$. The second
factor arises in a similar fashion from the contraction of the incoming
and outgoing polarization vectors with the $\epsilon$-tensor. The
possible polarization vectors for the incoming photon are
$\varepsilon_{1}=(0,1,0,0)$ and $\varepsilon_{2}=(0,0,1,0)$.
However, only the polarization vector $\varepsilon_{2}$ pointing
along the $y$-direction can occur for the kinematics we are considering;
only if the initial polarization is along the $y$-direction, will the
photon be deflected in the $xz$-plane. The reason is that $\varepsilon_{1}$
is (approximately) a linear combination of $q$ and $q'$
[as given by (\ref{eq-qprime})], and thus there
cannot be a nonvanishing contribution when all three vectors are
contracted with a common $\epsilon$-tensor.

For the outgoing photon, there are also two possible polarization vectors.
One of them is the unchanged, out-of-plane $\varepsilon_{2}$. However,
this vector is obviously impossible, since one factor of $\varepsilon_{2}$
must already be contracted with the $\epsilon$-tensor. Therefore, the
outgoing polarization must be $\varepsilon'_{1}\approx(0,\cos\theta,0,-\sin\theta)$,
which is the other unit vector perpendicular to $\vec{q}\,'$; and only the
$z$-component of this $\varepsilon_{1}'$
contributes to the contraction, providing the second factor of
$\sin\theta$ in ${\cal M}$.

For the conventional decay $\pi^{0}\rightarrow\gamma+\gamma$, there is an
analogous $\theta^{2}$ suppression factor in ${\cal M}$
when the angle between the decay photons is
small. The derivation of this factor follows along essentially the same lines
in that case. The fact that the small-angle behaviors of the matrix elements for
the two decays are similar will enable us to estimate the
$\gamma\rightarrow\gamma+\pi^{0}$ rate from the well-known
$\pi^{0}\rightarrow\gamma+\gamma$ rate.

Of course,
the rate for the novel Cerenkov-like process can be evaluated directly from the
matrix element and the Lorentz-violating kinematics. However, the kinematical part
of the calculation turns out to be extremely awkward. Instead, it is possible to
estimate the rate using what is known about the rate for the ordinary
two-photon decay of the $\pi^{0}$. This might initially seem problematic, since
the pion lifetime is normally calculated in the pion's rest frame, and the
parent photon in $\gamma\rightarrow\gamma+\pi^{0}$ has no equivalent rest frame.
However, an unstable particle's decay rate can certainly be calculated in a
frame in which the initial particle is moving, and we know that in a
Lorentz-invariant theory, the decay rate will be diminished (relative to the rest
frame rate) by the Lorentz factor $\gamma$.
Moreover, it is actually fairly straightforward
how this factor arises in the calculation. In the frame in which the parent
particle is moving, the decay products are preferentially emitted along the
direction of that motion. In an ultrarelativistic process, the decay products are
beamed into a narrow pencil of angles, of characteristic size $\sim m/E$
(where $m$ and $E$ are the mass and energy of the parent particle),
rather than covering the full $4\pi$ steradians. This limits the phase space
available for the decay, and the rate is reduced by precisely the time dilation
factor $\gamma^{-1}=m/E$.

The same kind of dilation occurs in the photon decays that may be allowed
in Lorentz-violating theories. While the parent particle in
$\gamma\rightarrow\gamma+\pi^{0}$ has no rest frame, the daughter particles
are overwhelmingly
bunched into an angular range $\sim m_{\pi}/E_{T}$. The characteristic lifetime
for a single particle decaying into two ultrarelativistic daughters, when the
interaction is governed by the anomaly term from (\ref{eq-L-A}), is
given by the $\pi^{0}$ lifetime of $\tau_{\pi}\approx 8.4\times 10^{-17}$ s.
So except very close to threshold, the rate of the $\gamma\rightarrow\gamma+\pi^{0}$
is $\Gamma\sim m_{\pi}/E_{T}\tau_{\pi}$.

The $\theta^{4}$ suppression of the $|{\cal M}|^{2}$ at small angles does not
befuddle this result, since the same suppression is present in the ordinary
$\pi^{0}\rightarrow\gamma+\gamma$ process.
However, there is a puzzling aspect to this.
With increasingly relativistic energies, the decay photons (in the $\pi^{0}$
decay) are beamed into a narrowing pencil of angles, but in spite of the
$\theta^{4}$ factor in $|{\cal M}|^{2}$, the rate decreases only quite
slowly. (The time dilation makes the rate
proportional to a single power of the separation angle $\theta$.)
Of course, the change in the decay rate with energy is entirely dictated
by Lorentz symmetry in this case. The resolution of this puzzle is that the
matrix element also grows with energy as $|{\cal M}|\propto E^{4}$, and
$E^{4}\theta^{4}\propto E^{4}(m_{\pi}/E)^{4}$ is independent of the energy.
The same characteristic argument applies in the Lorentz-violating Cerenkov-like
photon decay process.

If the transit time for a photon of energy $E>\alpha E_{T}$ from its
source to the Earth is much longer than $\sqrt{|\delta|}\tau_{\pi}$, the photon will
lose most of its energy through Cerenkov-like pion emission on its journey.
The factor $\alpha$ is present to ensure that the energy is not too close to
threshold; right above threshold, the pion emission angle falls to zero, and the
decay time differs significantly from $\sim\sqrt{|\delta|}\tau_{\pi}$. For the specific
observations we shall consider below, $\alpha=1.01$ turns out to be more than
sufficient.

Conversely, the observation of a photon from a
sufficiently distant source indicates that its
energy $E$ must be less than $\alpha E_{T}$; otherwise it would not have survived.
This places a bound on $\delta$:
\begin{equation}
\label{eq-boundformula}
\delta>-\frac{\alpha^{2}m_{\pi}^{2}}{2E^{2}}.
\end{equation}
Note that even for a 1 PeV photon, $E\tau_{\pi}/m_{\pi}\sim 10^{-9}$ s, so
effectively any astrophysical source is distant enough that the observation
of an emitted photon should produce a reliable bound. Since the $\gamma$-ray
spectrum of the Crab nebula extends up to at least 80 TeV~\cite{ref-aharonian12},
we can conclude that
\begin{equation}
\delta>-7\times 10^{-13}.
\end{equation}
This represents an improvement of more than an order of magnitude over the 
best previous bound.

The best previous constraint on a negative $\delta$ for the $\pi^{0}$ field
came from observations of $\gamma$-rays that were known to originate from
the $\pi^{0}\rightarrow\gamma+\gamma$ process. That was a major limitation,
since many of the highest-energy $\gamma$-rays appear to be produced by
inverse Compton scattering instead. The current method is much more widely
applicable. The survival of any $\gamma$-ray, regardless of how it originated,
over long astrophysical distances, allows us to place a useful bound.
The bound derived from this technique is the strongest presently available, since
it is be based on the very highest energy $\gamma$-ray observations. In the
not-so-distant future, it may be possible to observe photons with PeV energies
(since there is 
already evidence of individual particles with PeV energies in the
Crab nebula~\cite{ref-aharonian1});
that would lead to another two orders of magnitude improvement in the
constraint~(\ref{eq-boundformula}).

So the observed absence of photon decay processes gives a powerful way to
constrain Lorentz violation. However, this method does have a drawback. Any photon
decay process is generally only going to become allowed if the particles produced
in the decay have less energy than they would in the standard Lorentz-violating
theory. In other words, only negative values of $\delta$ can be constrained this
way. This also makes it difficult to study more general theories, in which the
Lorentz violation includes anisotropy as well as boost invariance violation.
The observation of photons coming from different directions would allow us to
place bounds on the $\delta(\hat{p})$ parameters corresponding to different
directions. However, with only one-sided bounds, it is not possible to disentangle
these bounds to get bounds on the individual $k^{\mu\nu}$ coefficients. The only
available bounds on a positive $\delta$ for the $\pi^{0}$ are at the $2\times 10^{-9}$
level, and this limits how tightly the individual $k^{\mu\nu}$ parameters may
be constrained.

Ultimately, the best
future constraints on Lorentz violation in the $\pi^{0}$ sector may
come from a clearer understanding of the relationships between the fundamental SME
coefficients for the quark and gluon fields and the coefficients for composite fields
like the pion. However, at present, the best bounds on dimension-four Lorentz-violating
operators for the $\pi^{0}$ field come from observations of TeV $\gamma$-rays. In this
paper, we have given an improved one-sided constraint, derived from the observed
absence of the Cerenkov-like emission process $\gamma\rightarrow\gamma+\pi^{0}$ for
Crab nebula $\gamma$-rays with up to 80 TeV energies. This represents an order of
magnitude improvement over the best previous bounds.

\end{document}